# The orbital period of the eclipsing dwarf nova SDSS J081610.84+453010.2

Jeremy Shears, Steve Brady, Tut Campbell, Arne Henden, Enrique de Miguel, Etienne Morelle, George Roberts, Richard Sabo and Ian Miller

## Abstract

We present time resolved photometry of the cataclysmic variable SDSS J081610.84+453010.2 and have established for the first time that it is an eclipsing dwarf nova. We observed an outburst of the system which lasted about 11 days and had an amplitude of 3.4 magnitudes above mean quiescence. From an analysis of the eclipse times of minimum during the outburst, we determined the orbital period as $P_{orb}$ = 0.2096(4) d or 5.030(10) h. The orbital period places it above the period gap in the distribution of orbital periods of dwarf novae. The eclipses are of short duration (average FWHM = 10.7 min or 0.036 of the orbital period) and shallow (average 0.4 mag during outburst and 0.6 magnitude in quiescence), suggesting a grazing eclipse.

## Background

Dwarf novae are a class of cataclysmic variable star in which a white dwarf primary accretes material from a secondary star via Roche lobe overflow. The secondary is usually a late-type main-sequence star. In the absence of a significant white dwarf magnetic field, material from the secondary passes through an accretion disc before settling on the surface of the white dwarf. As material builds up in the disc, a thermal instability is triggered that drives the disc into a hotter, brighter state causing an outburst in which the star apparently brightens by several magnitudes (1) (2).

SDSS J081610.84+453010.2 was first identified (3) spectroscopically as a cataclysmic variable in the first data release from the Sloan Digital Sky Survey (SDSS). SDSS lists the object as having g = 20.08. The spectrum showed evidence of an M-type secondary by the presence of a TiO band at 7100 Å, indicating a relatively low contribution of the accretion disk and, hence, a low mass transfer rate system. The star is located at RA 08h 16m 10.84s, Dec. +45º 30' 10.2" (J2000) in Lynx.

The AAVSO International Database lists 416 observations of SDSS J081610.84+453010.2 between 2002 February and the 2010 outburst discussed in this paper, but only 28 were positive detections. The majority of these were in the magnitude range of 18 to 20, i.e. near to, or slightly above, quiescence. There is only one other outburst recorded when the star was detected by Eddy Muyllaert at C = 16.9 (C, clear filter) on 2007 June 4. Two nights later it reached C= 16.1, but by the following day it was back near to quiescence (C = 19.4). Thus the outburst lasted only about 3 days. The faintness and short duration of this outburst suggests that others might have been missed. We investigated whether other outbursts might have been observed by the Catalina Real-Time Transient Survey (4) (CRTS), but unfortunately the star is at the edge of the relevant field and thus does not receive significant coverage (5).

In this paper we present photometry obtained in quiescence in 2002/3 and during the outburst of SDSS J081610.84+453010.2 in December 2010.

**Photometry and analysis**

Unfiltered photometry was conducted during the 2010 outburst and during quiescence using the instrumentation shown in Table 1 and according to the observation log in Table 2. Images were dark-subtracted and flat-fielded prior to being measured using differential aperture photometry relative to the AAVSO V-band sequence 3540opa (6). Given that each observer used slightly different instrumentation, including CCD cameras with different spectral responses, small systematic differences are likely to exist between observers. Where overlapping datasets were obtained during the outburst, we aligned measurements by different observers by experiment. Adjustments of up to 0.07 magnitudes were made. Heliocentric corrections were applied to all data.

**Outburst light curve**

The overall light curve of the outburst is shown in Figure 1 and expanded plots of some of the longer photometry runs are shown in Figure 2. The outburst of this cataclysmic variable confirms that it is a dwarf nova. The outburst was detected on 25 Dec 2010 (JD 2455555.791) when the star was brightening. Maximum brightness occurred later that day at magnitude 15.8, after which the star faded at an average rate of 0.30 mag/d over the next 2.5 days. Some 6 days after the outburst was detected the star was at magnitude 17.8, reaching magnitude 18.8 about 11 days after detection. Taking the average quiescence magnitude outside eclipse as V=19.2 (discussed below), means that the outburst amplitude was 3.4 magnitudes and the duration was about 11 days. It was therefore clearly different from the 2007 outburst. The light curve shows that dips are clearly present which, because of their periodicity, we interpret as eclipses.

**Detection of eclipses and measurement of the orbital period**

A total of 7 different eclipses were observed during the outburst. The times of minimum eclipse were determined using the Kwee and van Woerden (7) method in the Minima v2.3 software (8). We obtained 10 times of minimum as some eclipses were observed by two observers. The measured times of minimum are listed in Table 3 where the eclipses are labelled with the corresponding orbit number starting from 0. The orbital period was then calculated by an unweighted linear fit to these times of minimum as $P_{orb}$ = 0.2096(4) d or 5.030(10) h. The orbital period places SDSS J081610.84+453010.2 above the period gap in the orbital period of dwarf novae. This, coupled with the fact that the star undergoes infrequent outbursts, suggests it is a member of the SU UMa family of dwarf novae.

The eclipse time of minimum ephemeris given by the linear fit is:

$$HJD_{min} = 2455555.8512(6) + 0.2096(4)*E$$

The O-C (Observed – Calculated) residuals of the eclipse minima relative to this ephemeris are shown in Table 3. We also plot the residuals, given as a fraction of the orbital period, in the top panel of Figure 3.

We also observed 3 eclipses during quiescence in 2002 and 2003 (Figure 4) and their times of minimum are listed in Table 3. In each case the average magnitude outside eclipse was V ≈ 19.2. In eclipse the star was very faint at V ≈ 20 which was at the limit of the instrumentation and the eclipses are defined by rather few data points. We tried to obtain a linear fit to all the data, i.e. from quiescence and from the recent outburst, but it was not possible to find an unambiguous solution due to the relatively short time over which the 2010 eclipses were observed and the associated uncertainty in $P_{orb}$.

**Analysis of the eclipses**

An expanded view of some of the longer photometry runs during the outburst is presented in Figure 2, showing 5 of the eclipses, and in Figure 4 in quiescence. We measured the full width at half minimum (FWHM) duration of each eclipse relative to a baseline corresponding to the average light curve before and after the eclipse. In some cases, there were insufficient data points defining the eclipse and therefore no determination was possible. The eclipse FWHM data in Table 3 and the middle panel of Figure 3 show that the eclipse duration in outburst ranged between 8.5 min and 15.8 min with an average of 10.7 min. We must treat the range of FWHM values with caution as two problems were encountered in their measurement. Firstly, in several cases the minimum magnitude was likely not established precisely due to the low sampling rate of the photometry. Secondly, assigning the average baseline before and after the eclipse was complicated by the fact that the brightness outside eclipse was also varying. We note that the average FWHM eclipse duration of the only quiescence eclipse that could be measured with accuracy (Table 3) was the same as the average value during outburst, 10.7 min. This suggests that the eclipse duration, hence the diameter of the accretion disc is approximately the same regardless of the excitation level of the system. A similar behaviour has been seen in the SS Cyg-type dwarf nova CG Dra (9) and the SU UMa-type dwarf nova V713 Cep, during a normal outburst (10).

The average value of the eclipse duration, 10.7 min, represents a fraction of the orbital period of $\Delta\Phi_{1/2}$ = 0.036. The very short $\Delta\Phi_{1/2}$ suggests that the inclination, *i*, of the system is close to the critical value, below which eclipses do not occur, resulting in a grazing eclipse. The average critical value for dwarf novae is about $i = 71°$, but the limiting value for an individual system depends on the relative size of the secondary star (11). A well-known dwarf nova which undergoes grazing eclipses is U Gem. The system has $i = 77°$ (12) and $\Delta\Phi_{1/2}$ varies between 0.05 at quiescence and

0.11 at outburst (13). 1RXS J180834.7+101041, another dwarf nova has eclipses with $\Delta\Phi_{1/2}$ =0.064 and $i$ = 70$^o$ (14). Thus SDSS J081610.84+453010.2 may have an inclination similar to these dwarf novae.

The emission spectra of high inclination dwarf novae often show double-peaked emission lines due to the Doppler shift associated with the rotation of the accretion disc. However, the emission spectrum of SDSS J081610.84+453010.2 shown in Figure 5 is single-peaked. Although this might be due to the inclination being close to the critical value, this is not diagnostic as some high inclination systems have single-peaked spectra. This is particularly evident in longer orbital periods; one explanation is that as the period increases the accretion disc also increases in size, which in turns means that the material at the edge of the disc is moving more slowly. Other explanations have been invoked to explain the single peaks in high inclination systems. These usually rely on the primary emission source being something other than the rotating accretion disc. For example, a rotating polar accretion column associated with a white dwarf magnetic field has been suggested in some systems, e.g. in BT Mon (15). In the case of PG 1012-029 it has been proposed that the spectrum is dominated by a non-rotating bipolar wind from the accretion disc (16).

We also measured the depth of the eclipses, Δm, but as discussed above this was complicated by the low rate of data sampling during the eclipse which means that the minimum might have been missed which may result in an underestimate of Δm. The data in Table 3 and the bottom panel of Figure 3 show that during outburst Δm ≈ 0.3 to 0.6 magnitudes with an average of ~0.4 magnitudes. The three quiescence eclipses were somewhat deeper with Δm ≈ 0.7 to 0.9 magnitudes and an average of ~0.6 magnitudes. Such values of Δm are relatively modest compared to some eclipsing dwarf novae. For example, SDSS J150240.98+333423.9 (17) exhibits eclipses with depths up to 2.1 magnitudes and in the case of V713 Cep (10) up to 3 magnitudes. A modest Δm is consistent with a grazing eclipse in a system whose inclination is just above the critical value. U Gem has Δm = 0.02 to 0.9 magnitudes, depending on its excitation state (13).

Eclipse mapping and modelling in dwarf novae has been used to associate different parts of the eclipse profile with different parts of the system undergoing eclipse, including the white dwarf, the bright spot and the accretion disc. However, the time cadence was not sufficient to resolve any features in the eclipses of SDSS J081610.84+453010.2. In any case, if, as proposed, the system is just above the critical inclination, it is unlikely that the white dwarf itself would be eclipsed. Instead it is most likely that there is simply a partial eclipse of the accretion disc.

It is also apparent that the brightness was varying outside eclipse, with evidence of an orbital hump plus flickering which is seen in many dwarf novae and which is caused by variations in the mass transfer rate of material onto the accretion disc or the white dwarf (18). The orbital hump is rather unusual, because it occurs after the eclipse. By contrast, in most eclipsing dwarf novae the hump occurs before the eclipse and is due to an eclipse of the bright spot where the accretion stream hits the disc. This may suggest that in the case of SDSS J081610.84+453010.2 the hump is caused by a different phenomenon. Post-eclipse humps have been seen in the AM Her-type cataclysmic variable (or "polar") WW Hor (where pre-eclipse humps are also present) (19). Whilst there is no suggestion that SDSS J081610.84+453010.2 is a polar, on the contrary it appears to have an accretion disc since it undergoes outbursts, further studies of this intriguing post-eclipse hump would be helpful.

**Future outbursts**

The 2010 outburst reported here is the first well-observed outburst of SDSS J081610.84+453010.2 and, after the shorter one in 2007, is only the second on record. Whilst it is likely that other outbursts have been missed, due to incomplete observational coverage, the low accretion rate proposed for the system (3) suggests that outbursts may be infrequent. Therefore we encourage further monitoring with the aim of determining the outburst frequency. Photometry during future outbursts would also help to improve the accuracy of the orbital period and eclipse ephemeris. It will also be important to investigate the nature of the unusual orbital hump which we report in the 2010 outburst. The star was added to the BAA Variable Star Section's Recurrent Objects Programme in 2007 November. This programme was set up as a joint project between the BAA-VSS and *The Astronomer* magazine specifically to monitor poorly studied eruptive stars of various types where outbursts occur at intervals of greater than 1 year (20).

**Conclusions**

We have established for the first time that the SDSS J081610.84+453010.2 is an eclipsing dwarf nova. We observed an outburst of the system which lasted about 11 days with an amplitude of 3.4 magnitudes above mean quiescence. From an analysis of the eclipse times of minimum during the outburst, we determined the orbital period as $P_{orb}$ = 0.2096(4) d or 5.030(10) h. The eclipses are of short duration (average FWHM = 10.7 min, $\Delta\Phi_{1/2}$ = 0.036) and shallow (average 0.4 mag during outburst and 0.6 magnitude in quiescence), which suggests these are grazing eclipses. An orbital hump was observed which, by contrast to the situation in most dwarf novae, occurred just after the eclipse.

**Acknowledgements**

The authors gratefully acknowledge the use of observations from the AAVSO International Database contributed by observers worldwide. This research made use of the Sloan Digital Sky Survey (SDSS) and the NASA/Smithsonian Astrophysics Data System. We thank the referees, Dr Chris Lloyd and Dr Robert Connon Smith, for helpful comments which have improved the paper.

**Addresses**

JS: "Pemberton", School Lane, Bunbury, Tarporley, Cheshire, CW6 9NR, UK [bunburyobservatory@hotmail.com]
SB: 5 Melba Drive, Hudson, NH 03051, USA [sbrady10@verizon.net]
TC: 7021 Whispering Pine, Harrison, AR 72601, USA [jmontecamp@yahoo.com]
AH: AAVSO, 49 Bay State Rd. Cambridge, MA 02138 [arne@aavso.org]
EdM: Departamento de Fisica Aplicada, Facultad de Ciencias Experimentales, Universidad de Huelva, 21071 Huelva, Spain; Center for Backyard Astrophysics,

Observatorio del CIECEM, Parque Dunar, Matalascañas, 21760 Almonte, Huelva, Spain [demiguel@uhu.es]
EM: Lauwin-Planque Observatory, F-59553 Lauwin-Planque, France [etmor@free.fr]
GR: 2007 Cedarmont Dr., Franklin, TN 37067, USA, [georgeroberts@comcast.net]
RS: 2336 Trailcrest Dr., Bozeman, MT 59718, USA [richard@theglobal.net]
IM: Furzehill House, Ilston, Swansea, SA2 7LE, UK [furzehillobservatory@hotmail.com]

| Observer | Telescope | CCD |
|---|---|---|
| Brady | 0.4 m reflector | SBIG ST-8XME |
| Campbell | 0.2 m SCT | SBIG ST-7 |
| Henden | 1.0m reflector [a] | 1024x1024 SITe/Tektronix thinned, back-illuminated |
| de Miguel | 0.25 m reflector | QSI-516ws |
| Miller | 0.35 m SCT | Starlight Xpress SXVF-H16 |
| Morelle | 0.4 m SCT | SBIG ST-9 |
| Roberts | 0.4 m SCT | SBIG ST-8 |
| Sabo | 0.43 m reflector | SBIG STL-1001 |
| Shears | 0.28 m SCT | Starlight Xpress SXVF-H9 |

[a] U.S. Naval Observatory, Flagstaff Station

**Table 1: Instrumentation**

| Start date UT | Start time HJD | End time HJD | Duration (h) | Observer |
|---|---|---|---|---|
| *Quiescence data* | | | | |
| 2002 Feb 12 | 2452317.609 | 2452317.965 | 8.5 | Henden |
| 2003 Apr 22 | 2452751.621 | 2452751.726 | 2.5 | Henden |
| 2003 Apr 28 | 2452757.629 | 2452757.766 | 3.3 | Henden |
| *Outburst data* | | | | |
| 2010 Dec 25 | 2455555.790 | 2455555.951 | 3.9 | Brady |
| 2010 Dec 25 | 2455556.367 | 2455556.541 | 4.2 | Shears |
| 2010 Dec 26 | 2455557.348 | 2455557.625 | 6.6 | Morelle |
| 2010 Dec 26 | 2455557.429 | 2455557.740 | 7.5 | de Miguel |
| 2010 Dec 27 | 2455557.649 | 2455558.058 | 9.8 | Sabo |
| 2010 Dec 27 | 2455557.683 | 2455558.015 | 8.0 | Campbell |
| 2010 Dec 28 | 2455558.638 | 2455559.003 | 8.8 | Roberts |
| 2010 Dec 30 | 2455561.361 | 2455561.365 | 0.1 | Miller |
| 2011 Jan  3 | 2455565.423 | 2455565.433 | 0.2 | Miller |

**Table 2: Observation log**

| Time of minimum (HJD) | Error (d) | Orbit number | O-C (d) | Eclipse duration FWHM (min) | Eclipse depth (mag) |
|---|---|---|---|---|---|
| *Quiescence data* | | | | | |
| 2452317.7988 | 0.0035 | | | ND | 0.72 |
| 2452751.6352 | 0.0035 | | | 10.7 | 0.88 |
| 2452757.7231 | 0.0030 | | | ND | 0.71 |
| *Outburst data* | | | | | |
| 2455555.8495 | 0.0018 | 0 | -0.0017 | ND | 0.32 |
| 2455556.4811 | 0.0040 | 3 | 0.0013 | 10.1 | 0.49 |
| 2455557.5207 | 0.0009 | 8 | -0.0069 | 11.4 | 0.32 |
| 2455557.5294 | 0.0010 | 8 | 0.0019 | 11.6 | 0.33 |
| 2455557.7389 | 0.0009 | 9 | 0.0017 | 15.8 | 0.63 |
| 2455557.7393 | 0.0005 | 9 | 0.0022 | 8.5 | 0.45 |
| 2455557.9489 | 0.0008 | 10 | 0.0023 | 11.4 | 0.33 |
| 2455557.9493 | 0.0010 | 10 | 0.0027 | 8.5 | 0.29 |
| 2455558.7835 | 0.0007 | 14 | -0.0014 | 10.7 | 0.38 |
| 2455558.9928 | 0.0011 | 15 | -0.0016 | 8.5 | 0.41 |

**Table 3: Eclipse times of minimum, duration and depth**

The first 10 eclipses were observed during the 2010 outburst and the last 3 during quiescence in 2002 and 2003. ND: not determined

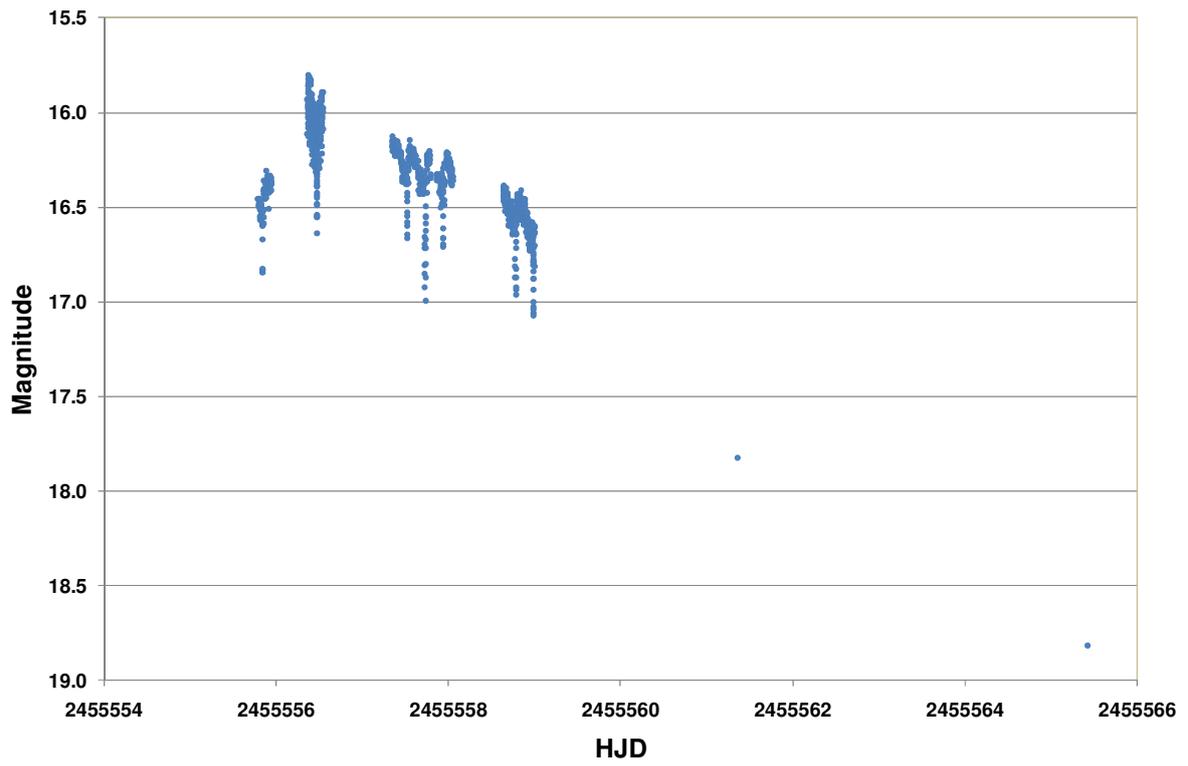

**Figure 1: Outburst light curve of SDSS J081610.84+453010.2**

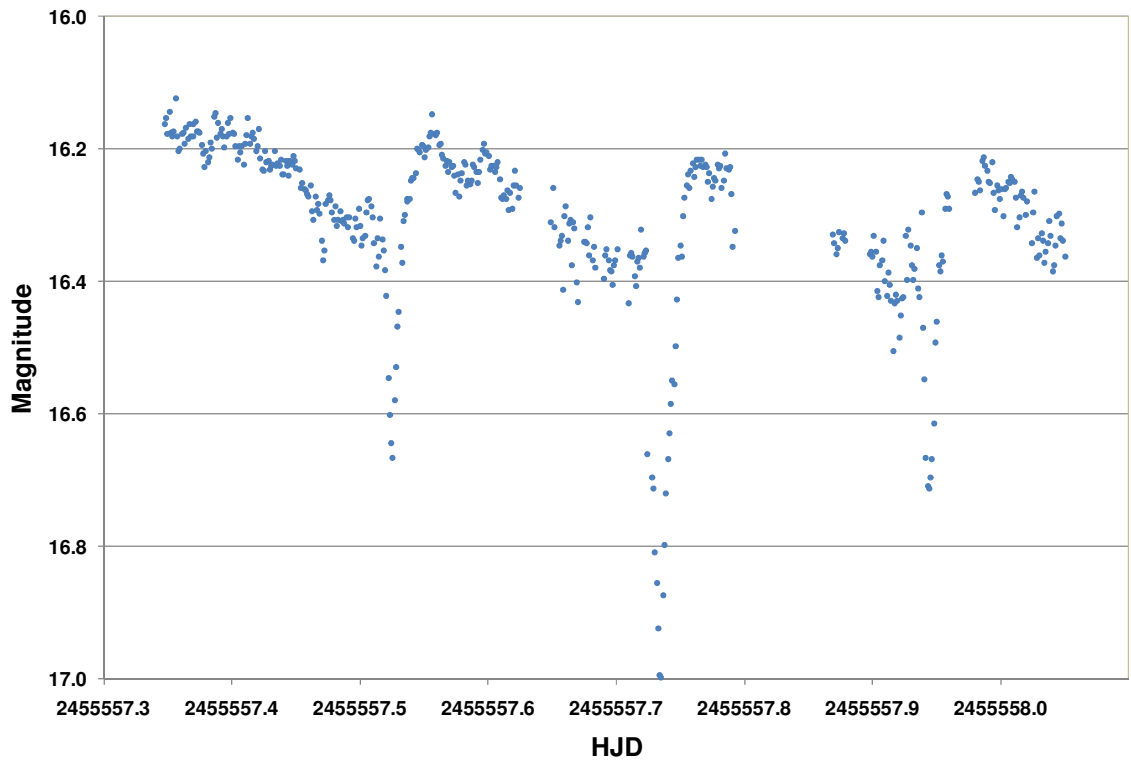

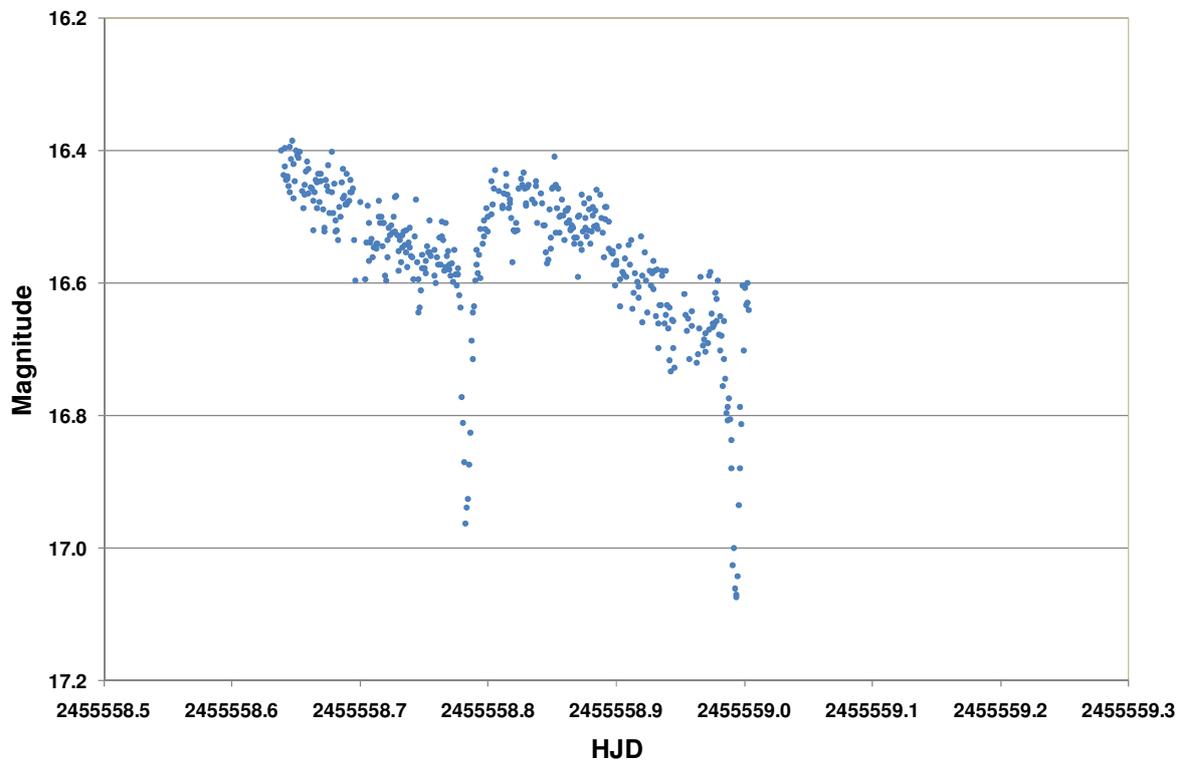

**Figure 2: Time resolved photometry showing eclipses during the outburst**

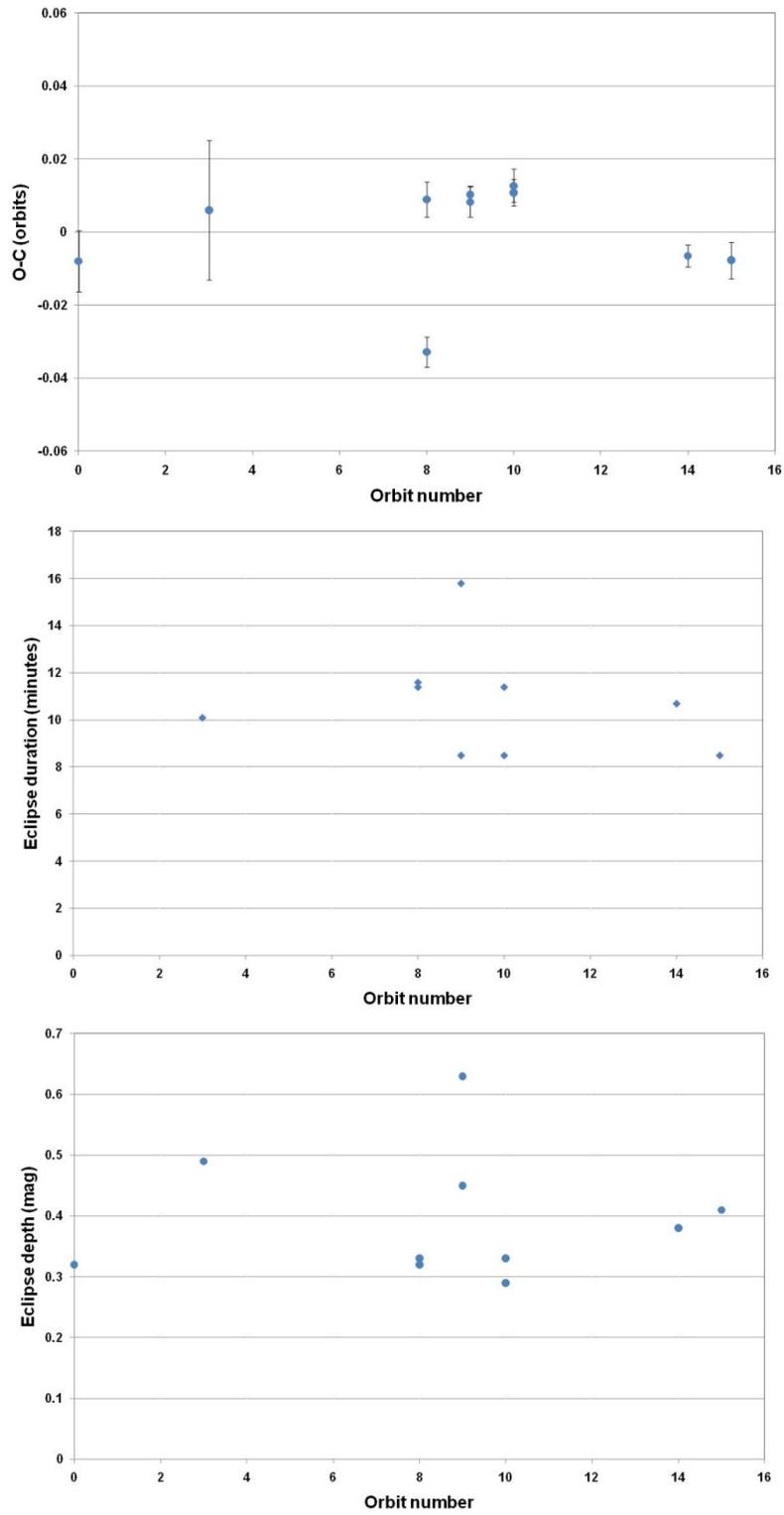

**Figure 3: (top) O-C for the eclipses, (middle) eclipse duration, (bottom) eclipse depth**

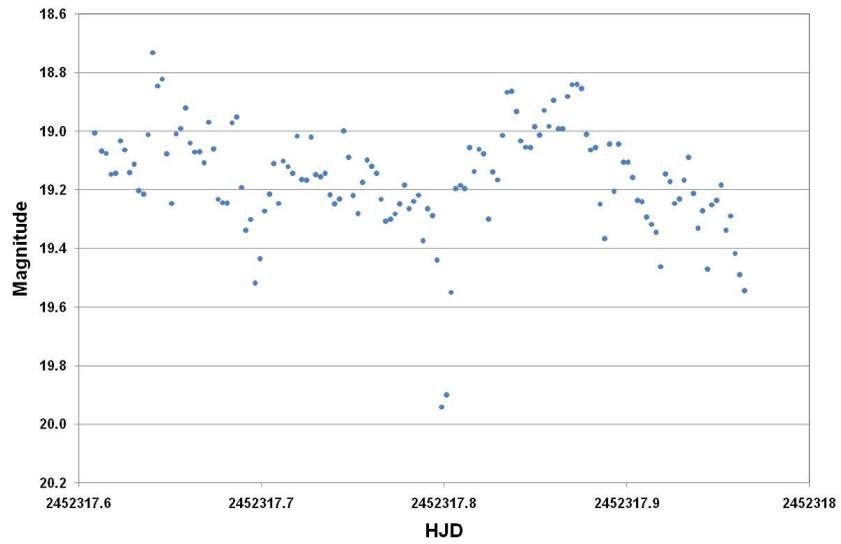

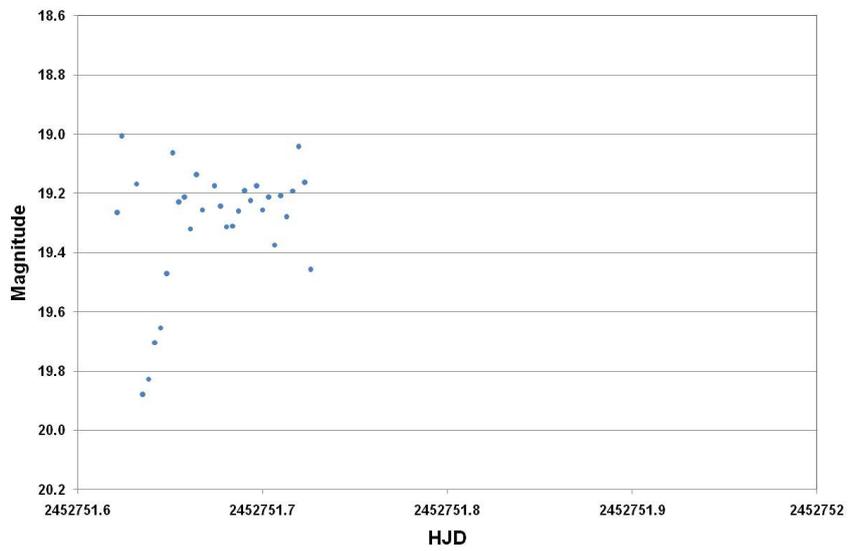

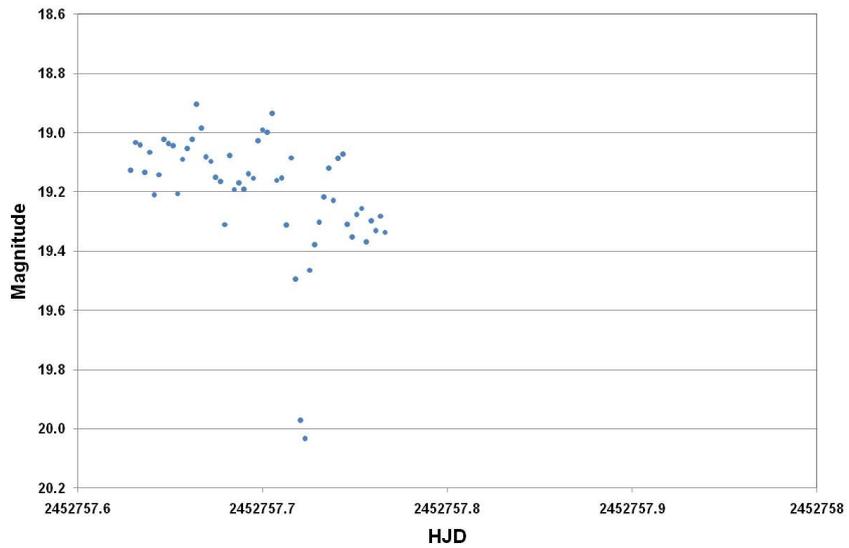

**Figure 4: Eclipses at quiescence**

(top) 2002 Feb 12, (middle) 2003 Apr 22, (bottom) 2003 Apr 28

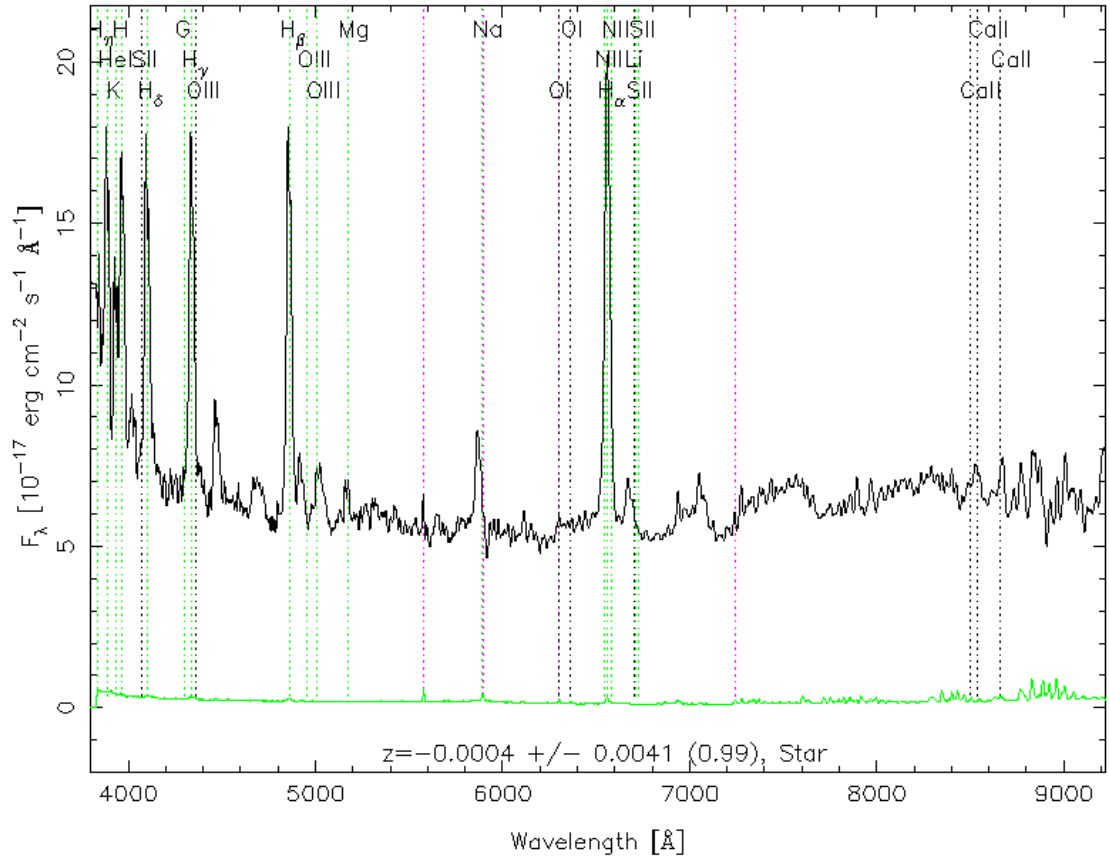

**Figure 5: Spectrum of SDSS J081610.84+453010.2**

Data from SDSS DR7 (21)